\documentclass[pra,aps, twocolumn, nobalancelastpage, superscriptaddress, nofootinbib]{revtex4-1}
\usepackage{mathptmx}

\usepackage{amssymb}
\usepackage{amsmath}
\usepackage{mathtools}
\usepackage{mathrsfs}
\usepackage{graphicx}
\usepackage{amsfonts}
\usepackage{color}
\usepackage{bm}
\usepackage{xspace,soul}
\usepackage{hyperref}
\usepackage{comment}
\usepackage{tikz}
\usetikzlibrary{calc,decorations.markings}
\makeatletter
\def\clearemail{\let\@FMN@list\@empty}
\makeatother

\DeclarePairedDelimiter{\abs}{\lvert}{\rvert}
\usepackage{braket}
\newcommand{\nepero}{\text{e}}
\newcommand{\Li}[1]{\mathrm{Li}_{#1}\!}
\newcommand{\de}{\mathrm{d}}

\newcommand{\Tr}{\mathrm{Tr}\,}

\renewcommand{\Im}{\mathop{\text{Im}}\nolimits}%
\newcommand{\ii}{\mathrm{i}}
\usepackage[normalem]{ulem}

\graphicspath{{images/}}

\begin{document}

\title{Algebraic Localization from Power-Law Interactions in Disordered Quantum Wires}
\author{Thomas Botzung}
\affiliation{University of Strasbourg, CNRS, ISIS (UMR 7006) and IPCMS (UMR 7504), 67000 Strasbourg, France}
\affiliation{Dipartimento di Fisica e Astronomia dell'Universit\`a di Bologna, I-40127 Bologna, Italy}
\author{Davide Vodola}
\email{davide.vodola@gmail.com}
\affiliation{Department  of  Physics,  Swansea  University,  Singleton  Park,  Swansea  SA2  8PP,  United  Kingdom}
\author{Piero Naldesi}
\affiliation{Universit\'{e} Grenoble-Alpes, LPMMC, F-38000 Grenoble, France \\ and CNRS, LPMMC, F-38000 Grenoble, France}
\author{Markus M\"uller}
\affiliation{Department  of  Physics,  Swansea  University,  Singleton  Park,  Swansea  SA2  8PP,  United  Kingdom}
\author{Elisa Ercolessi}
\affiliation{Dipartimento di Fisica e Astronomia dell'Universit\`a di Bologna, I-40127 Bologna, Italy}
\affiliation{INFN, Sezione di Bologna, I-40127 Bologna, Italy}
\author{Guido Pupillo}
\email{pupillo@unistra.fr}
\affiliation{University of Strasbourg, CNRS, ISIS (UMR 7006) and IPCMS (UMR 7504), 67000 Strasbourg, France}

\begin{abstract}
We analyze the effects of disorder on the correlation functions of one-dimensional quantum models of fermions and spins with long-range interactions that decay with distance $\ell$ as a power-law $1/\ell^\alpha$. Using a combination of analytical and numerical results, we demonstrate that power-law interactions imply a long-distance algebraic decay of correlations within disordered-localized phases, for all exponents $\alpha$. The exponent of algebraic decay depends only on $\alpha$, and not, e.g., on the strength of disorder. We find a similar algebraic localization for wave-functions. 
These results are in contrast to expectations from short-range models and are of direct relevance for a variety of quantum mechanical systems in atomic, molecular and solid-state physics. 
\end{abstract}
\maketitle

Quantum waves are generally localized exponentially by disorder. Following the seminal work by Anderson with spin-polarized electrons~\cite{Anderson1958}  much experimental~\cite{Roati2008, Billy2008, Kondov2011, Jendrzejewski2012, Schreiber2015, Smith2016, Bordia2016} and theoretical interest has been devoted to the study of localized phases and to the localization-delocalization transition for non-interacting and interacting quantum models~\cite{Lee1985, Altshuler1997, Basko2006, Oganesyan2007, Gornyi2005, Znidaric2008,  Biddle2009, Biddle2010, Biddle2011,  Pal2010, DeLuca2013, Vosk2013, BarLev2014, BarLev2015, Luitz2015, Luitz2016, Naldesi2016, Nandkishore2014, Gornyi2017, Potter2016}. 
 
While most works have focused on short-range couplings, long-range hopping and interactions that decay with distance~$\ell$ as a power-law $1/\ell^{\alpha}$ have recently attracted significant interest~\cite{Kastner2010, Gong2014, FossFeig2015, Hauke2013, Eisert2013, Schachenmayer2013, Metivier2014, Kastner2015, Cevolani2015,Cevolani2016, Bettles2017, Frerot2017,Frerot2018} as they  can be now engineered in a variety of atomic, molecular and optical systems. For example, power-law spin interactions with tunable exponent $0<\alpha<3$ can be realized in arrays of laser-driven cold ions~\cite{Schneider2012a,Richerme2014, Jurcevic2014, Britton2012,Bermudez2013} or between atoms trapped in a photonic crystal waveguide~\cite{Shahmoon2013,Douglas2015,Litinskaya2016,Vaidya2018}; dipolar-type $1/\ell^{3}$ or van-der-Waals-type $1/\ell^{6}$ couplings have been experimentally demonstrated with ground-state neutral atoms~\cite{Kadau2016, Lepoutre2018, Baier2018, Tang2018}, Rydberg atoms~\cite{Weimer2008,Saffman2010,  Viteau2012, Schauss2012,Carr2013, Barredo2014, Balewski2014, Jau2015, Weber2015,Faoro2016,Labuhn2016,Gorniaczyk2016, Zeiher2016, Bernien2017, Pineiro2018}, polar molecules~\cite{Yan2013, Hazzard2014,Reichsollner2017} and nuclear spins~\cite{Alvarez2015}. In solid state materials, power-law hopping is of interest for, e.g., excitonic materials~\cite{Anderson1998, Anderson2002, Scholes2006, Dubin2005, Dubin2006, Vogele2009, Wuster2011,Gunter2013,  Robicheaux2014, Schonleber2015, Schempp2015, Barredo2015, Rosenberg2018}, while long-range $1/\ell$ coupling is found in helical Shiba chains~\cite{Pientka2013, Pientka2014}, made of magnetic impurities on an s-wave superconductor. 
In many of these systems, disorder - in particles' positions, local energies, or coupling strengths - is an intrinsic feature, and understanding its effects on single-particle and many-body localization remains a fundamental open question.

For non-interacting models, it is generally expected that long-range hopping induces delocalization in the presence of disorder for $\alpha< d$, while for $\alpha>d$ all wave-functions are exponentially localized~\cite{Anderson1958, Rodriguez2000, Rodriguez2003, DeMoura2005, Celardo2016}. 
However, recent theoretical works with positional~\cite{Deng2018} and diagonal~\cite{Celardo2016} disorder have demonstrated that localization can survive even for $\alpha<d$. Surprisingly, wave-functions were found to be localized only algebraically in these models, in contrast to the usual Anderson-type exponential localization expected from short-range models. How these finding translate to the behavior of wave-functions and, crucially, correlation functions in many-particle systems is not known.

In this work, we investigate the effects of disorder on the decay of correlation functions and wave-functions in long-range quantum wires of fermions and spins. These are extensions of the Kitaev chain with long-range pairing~\cite{Vodola2014,Vodola2016,Liu2018} and the Ising model in transverse field~\cite{Koffel2012}, corresponding to integrable and non-integable chains in the absence of disorder, respectively. For fermions, we determine the regimes of localization for all $\alpha$ for the cases of disordered hopping or pairing. For the Ising chain, we focus on the regime $\alpha>1$, where the disordered phase diagram has been shown to display many-body localization theoretically~\cite{Burin2015_2} and experimentally~\cite{Smith2016}. For all models we compute  the one-body and two-body connected correlation functions, finding several novel features: (i) The connected correlation functions decay algebraically at long distance within all localized phases, (ii) with an exponent that depends exclusively on $\alpha$, and not, e.g., on the disorder strength. (iii) For the fermionic models, we derive analytic results for the long-distance decay of the correlations that explain the found algebraic decay, in excellent agreement with the numerics. (iv) The same analytical predictions are found to hold also for the correlations of the interacting Ising chain. (v) For any $\alpha$, the localized wave-functions of the fermionic models display a long-distance algebraic decay with exponent $\alpha$, different from recent predictions for long-range hopping models. These results should be of direct relevance to many experiments in cold atomic, molecular and solid-state physics with fermions and spins.

\textit{Models.---} We consider the following Hamiltonians for one-dimensional long-range fermionic random models     
    \begin{equation}\label{eqn:kitaevHamiltonian}
    H_{{\rm I}, {\rm II}}   =  H_0 + V_{{\rm I}, {\rm II}}
    \end{equation}
where $H_0$ is a homogeneous Hamiltonian given by
\begin {equation}
\begin{split}
    H_0 &= - t \sum_{j=1}^{L} \left(a^\dagger_j a_{j+1} + \mathrm{H.c.}\right)  + \mu \sum_{j=1}^{L} n_j \\& +  \sum_{j,\ell} \frac{\Delta}{\ell^{\alpha}} \left( a_j a_{j+\ell}  + \mathrm{H.c.} \right) 
    \end{split}
    \end{equation}
that describes a $p$-wave superconductor with a long-range pairing, and the indices ${{\rm I}, {\rm II}}$ refer to the two different types of Hamiltonians we consider, namely   
    \begin{equation}
    V_{\rm I} = \sum_{j=1}^{L} W_j \left(a^\dagger_j a_{j+1} + \mathrm{H.c.}\right)
    \end{equation}
    that corresponds to a random hopping and
    \begin{equation}
    V_{\rm II} =\sum_{j,\ell} \frac{W_{j}}{\ell^{\alpha}} \left( a_j a_{j+\ell}  + \mathrm{H.c.} \right)
    \end{equation}
    that corresponds to a random long-range pairing.
In the previous equations, $a^{\dagger}_j$  $(a_j)$ is a fermionic creation (annihilation) operator on site $j$, $\mu$ is the chemical potential, $n_j = a^\dag_j a_j$ and $W_j$ are i.i.d random variables  drawn from a uniform distribution of width $2W$ and zero mean value.
 We fix the energy scale by letting $\Delta = 2t = 1$ and we choose $\mu = 2.5$, corresponding to a gapped paramagnetic phase for $W_j=0$~\cite{Vodola2014}. Different values of $\mu$ do not change the results we find in the following.  The random Hamiltonians~\eqref{eqn:kitaevHamiltonian} can be written in diagonal form as $H_{{\rm I}, {\rm II}} =\sum_{q=0}^{L-1} \Lambda_q \eta^\dag_q \eta_q $  by a generalized Bogoliubov transformation defined by $\eta_q = \sum_{j} ( g_{q,j} a_j + h_{q,j} a^\dag_j) $~\cite{Lieb1961}, with $\Lambda_q$ the energies of the single-particle states labelled by~$q$. The ground state $\ket{\Omega}$ is then the vacuum of all quasi-particles~$\eta_q$ and the matrix elements $g_{q,j}$ and $h_{q,j}$ can be identified with the wave functions of the two fermionic modes $\eta_{q}^\dag$ and $\eta_{q}$, respectively.

As an interacting model, we consider the following random long-range Ising model~\cite{Koffel2012} in transverse field 
\begin{equation}
H_\text{LRI} =  \sum_{j,\ell} {(\sin \theta + B_j)}\frac{\sigma^x_j\sigma^x_{j+\ell}}{\ell^\alpha} +  \sum_{j=1}^L (\cos\theta + W_j) \sigma^z_j,
\label{eqn_LRI}
\end{equation}
where $\sigma_j^\nu$ ($\nu=x,z$) are Pauli matrices for a spin-1/2 at site~$j$ and $B_j$  are i.i.d random variables  drawn from a uniform distribution of width $2B$ and zero mean value. We choose $\theta=\pi/5$, corresponding to a paramagnetic phase for $B_j = W_j = 0$~\cite{Vodola2016}. Different values of $\theta$ will not change the results we find in the following. 
For any finite disorder strength, the model Eq.~\eqref{eqn_LRI} has been shown to display a many-body localized (MBL) phase for $\alpha>2$~\cite{Hauke2015, Li2016, Maksymov2017}. 

In the following, we first determine the regimes of localization for the fermionic models Eqs.~\eqref{eqn:kitaevHamiltonian} and then compute the single- and two-body correlation functions, as well as the wave-functions, within the localized phases using a combination of analytical and numerical techniques. For the long-range Ising model Eq.~\eqref{eqn_LRI}  we compute the spin-spin connected correlation functions numerically. Our goal is to demonstrate that all these quantities decay algebraically at large distances both for non-interacting and interacting MBL localized models and to characterize their decay exponents. 

\begin{figure}
\includegraphics[width=\columnwidth]{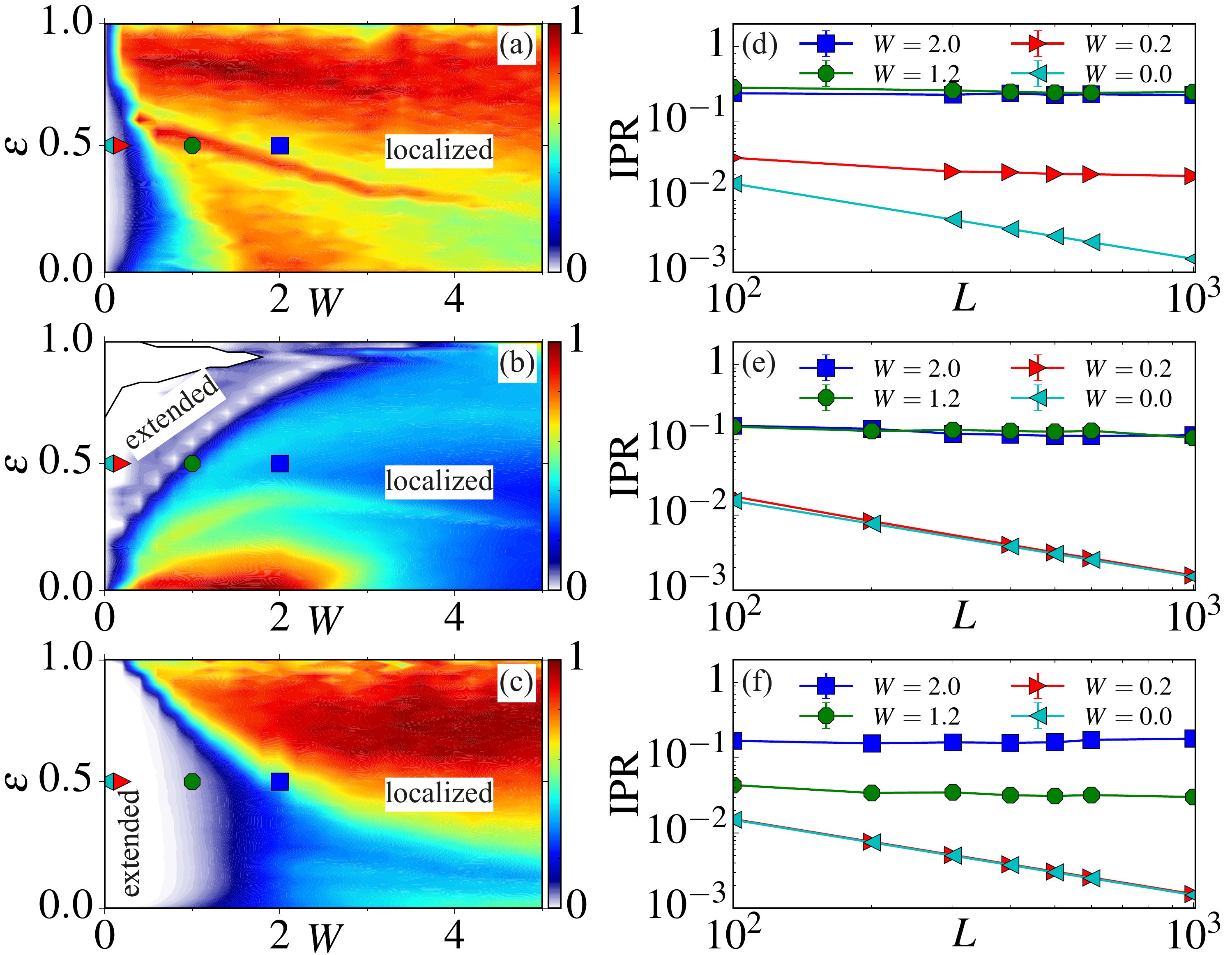}
\caption{Left panels: IPR in the thermodynamic limit as a function of the disorder strength $W$ and the rescaled energy~$\epsilon$ for (a) $\alpha=3$ and (b) $\alpha=0.8$ for the model (I) with random hopping and (c) for $\alpha=3.0$  for the model (II) with random pairing. In panel (b) the solid black line marks the region where the number of extended states is too low for a meaningful data analysis. Only in these panels, for drawing purpose, the IPR has been rescaled to 1 in correspondence of its maximum value. Right panels:  Scaling of the IPR as a function of the system size $L$ for $\epsilon=1/2$ and different $W$ for (d) $\alpha=3$ and (e) $\alpha=0.8$ for the model (I) and (f) for $\alpha=3.0$ for the model (II). In panels (a-c) the symbols indicate the values of $W$ and $\epsilon$ we choose to plot the IPR in panels (d-f).}\label{fig_IPR}
\end{figure}

\textit{Localized phases of disordered fermions.---} We determine the localized phases for Hamiltonians Eqs.~\eqref{eqn:kitaevHamiltonian} by combining information from the numerical calculation of the inverse participation ratio (IPR) and the entanglement entropy~\cite{supp_mat}. The IPR gives information about the spatial extension of single-particle states and is defined as $\mathrm{IPR}_q = {\sum_{j=1}^L[{\abs{g_{q,j}}^4 + \abs{h_{q, j}}^4}]}$ for a normalized state with energy~$\Lambda_q$. The IPR tends to zero for increasing $L$ for extended states, while it remains finite for localized states. If a value for the energies $\Lambda_q$ exists that separates extended states from localized states the system is said to display a (single-particle) mobility edge. 

For comparing the IPR of states with different energies, we rescale the $\Lambda_q$ (obtained for $\sim 200$ disorder realizations) according to $\epsilon_q=(\Lambda_q-\Lambda_\text{min})/(\Lambda_\text{max}-\Lambda_\text{min})$, with $\Lambda_\text{max}$ ($\Lambda_\text{min}$) the maximum (minimum) value of the energies $\Lambda_q$. We then bin the different levels into groups with equal energy width, we average the IPR within each bin. Finally, in order to obtain the phase diagrams, we perform a finite-size scaling of the obtained IPR in the limit $L\to\infty$~\cite{supp_mat}.

Figure~\ref{fig_IPR} shows exemplary results for IPR as a function of $W$ and $\epsilon$ for model Eqs.~\eqref{eqn:kitaevHamiltonian} (I) [for $\alpha=3$ and 0.8 in panels (a) and (b), respectively], and (II) [for $\alpha=3$ in panel (c)] together with examples of finite size scaling [panels (d-e)]. The figure shows that the phase diagrams are much richer than expected from pure long-range hopping models: For model (I) with disordered hopping and $\alpha>1$ [panel (a)] essentially all states are localized. For $\alpha<1$ [panel (b)]  we find that, at $W$ fixed, there exists a mobility edge below (above) which all the states are localized (delocalized). For  model (II) [panel (c)]  with disordered pairing when $\alpha>1$ localized states are present at all energies if $W\gtrsim 2$, while we find a mobility  edge for $\alpha>1$ and $W\lesssim 2$: all states are delocalized at low energy $\epsilon$ and localized for higher $\epsilon$. Below we focus on the identified localized phases and compute the correlation functions and wave-functions for all models. 

\textit{Correlation functions.---} We consider the single-particle correlator $C(j,\ell) = \braket{a^\dag_{j} a_{j+\ell} }_W$  for the two free-fermionic models of Eqs.~(\ref{eqn:kitaevHamiltonian}) as well as the spin-spin correlation function $S_{\nu}(j,\ell) = \left[\braket{\sigma^{\nu}_{j} \sigma^{\nu}_{j+\ell}}- \braket{\sigma^{\nu}_{j}} \braket{\sigma^{\nu}_{j+\ell}}\right]_W $ (for $\nu=x,z$) for the interacting long-range Ising model of Eq.~(\ref{eqn_LRI}). In the definitions of $C(j,\ell)$ and $S_{\nu}(j,\ell)$  the subscript $W$ indicates averaging over the disorder distribution. For models with short-range interactions, all the correlation functions decay exponentially with~$\ell$. Here we are interested in the effects of long-range interactions.

\begin{figure}
\includegraphics[width=\columnwidth]{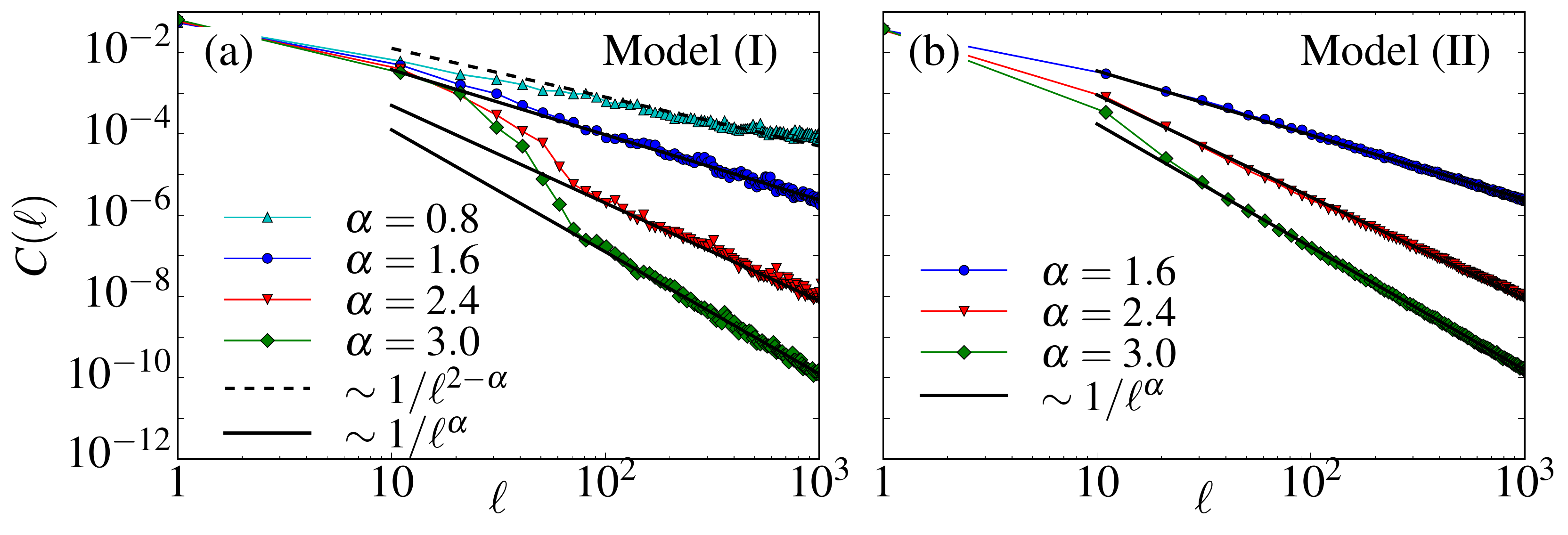} 
\caption{(a) Correlation function $C(\ell)$ for the model (I) as a function of the lattice site $\ell$ for different values of $\alpha$ and for $W=5$, $L=2000$ and 400 disorder realizations. The power-law tails are fit by the black lines scaling as $1/\ell^{2-\alpha}$ (dashed) and $1/\ell^{\alpha}$ (solid) in agreement with the analytical results in Eq.~\eqref{eqn_correctionGreenFunction}.  (b) Same as panel (a) but for the model (II).}\label{fig_CorrelationFunctions}
\end{figure}

Figures~\ref{fig_CorrelationFunctions}(a) and (b) show representative results for the correlator $C(\ell) := C(j_0,\ell)$ for models $H_\text{I}$ and $H_\text{II}$, respectively, for different values of $\alpha$. We choose $j_0 = L/4$ far from the edges in order to avoid boundary effects. We find numerically that the long-distance decay of $C(\ell)$ is always of power-law type $C(\ell)\sim \ell^{-\gamma}$ for all $\alpha$ within localized phases. In particular, for model $H_\text{I}$ [panel (a)] and $\alpha<1$ the decay is essentially algebraic at all distances with $\gamma \sim {2-\alpha}$, while for $\alpha>1$ we find for both models a hybrid decay that is exponential at short distances and power-law at large distances, with $\gamma \sim {\alpha}$ [panels (a) and (b)].  
Remarkably, we find that the values of the decay exponents of the power-law tails  do not depend on the disorder strength $W$~\cite{supp_mat}. 

This surprising long-distance behavior of correlations can be understood by computing the correlations analytically treating disorder as a perturbation. Here, we focus on model (I) with perturbation $ V_{\rm I}$, while a similar argument can be applied also to (II).  
The homogeneous Hamiltonian $H_0$ can be diagonalised via Fourier and Bogoliubov transformations as  $H_0 = \sum_k \lambda_{\alpha}(k) \xi^\dag_k \xi_k$, where  $\lambda_{\alpha}(k) = [\left( \cos k - \mu \right)^2 + 4f_{\alpha}^2(k)]^{1/2}$ and $\xi_k$ are extended Bogolioubov quasi-particles related to the unperturbed fermionic operators in momentum space via $\tilde{a}_k = v_k  \xi_k - u_k  \xi^\dag_{-k}$ with $v_k= \cos\varphi(k)$ and $u_k= \ii \sin\varphi(k)$, with $\tan 2\varphi(k) =  f_{\alpha}(k) / [\mu - \cos k]$ and $f_{\alpha}(k) = \sum_{\ell=1}^{L-1} \sin(k \ell)/\ell^\alpha$~\footnote{The functions $f_{\alpha}(k)$ when $L\to\infty$ become $f_{\alpha}(k) =  \left[\Li{\alpha}(\nepero^{\ii k})-\Li{\alpha}(\nepero^{-\ii k})\right]/(2\ii)$, with $\Li{\alpha}(z) = \sum_{j}z^j/j^\alpha$ a polylogarithm of order $\alpha$~\cite{Abramowitz1964}}.
At first order in $W_j$ the ground state $\ket{\Omega_0}$ of the unperturbed Hamiltonian $H_0$ is modified by $ V_{\rm I} $ as
\begin{equation}\label{eqn:PerturbedGSHopping}
\ket{\Omega} =\ket{\Omega_0} + \ket{\delta\Omega_0}=\ket{\Omega_0} -  \sum_{kk'} {J}_{k,k'} A(k,k')  \xi^\dag_k \xi^\dag_{k'}\ket{\Omega_0},
\end{equation}
where we define $J_{k,k'} = -\sum_j e^{\ii (k-k') j} W_j/L$ and $A(k,k') = 2(\nepero^{\ii k} + \nepero^{-\ii k'}) v_k u^*_{k'} / [\lambda(k)+\lambda(k')]$. 
Since $\braket{J_{k,k'}}_W = 0$, we note that the terms $\braket{\delta\Omega_0  | a^\dag_j a_{j+\ell} |  \Omega_0}_W$ and $\braket{\Omega_0| a^\dag_j a_{j+\ell} | \delta\Omega_0}_W$ vanish due to averaging over the disorder distribution. Thus, we obtain the following expression for $C(\ell)$ 
\begin{equation}\label{eqn_correlator_perturbation}
\braket{\Omega | a^\dag_j a_{j+\ell} | \Omega }_W =  
\braket{\Omega_0    | a^\dag_j a_{j+\ell} |      \Omega_0} +
\braket{\delta\Omega_0 	| a^\dag_j a_{j+\ell} |  \delta\Omega_0}_W.
\end{equation}

The first term in the r.h.s.~of Eq.~\eqref{eqn_correlator_perturbation} corresponds to the correlator of the homogeneous system~\cite{Vodola2016} that is $\braket{\Omega_0 	| a^\dag_j a_{j+\ell} |\Omega_0} = \int_{0}^{2\pi}\de k \ \nepero^{\ii k \ell} R_0(k)$, with $R_0(k) = \abs{u_k}^2$. The second term arises instead because of the random part of the Hamiltonian and reads 
\begin{equation}\label{eqn_IntegralRandomCorr}
\braket{\delta\Omega_0 	| a^\dag_j a_{j + \ell} |  \delta\Omega_0}_W = \frac{2W^2}{3}\int_{0}^{2\pi} \de k   \ \nepero^{\ii k \ell} R_1(k),
\end{equation}
where we have defined $R_1(k)= [c - {U}(k)] \abs{u_k}^2 -  {V}(k) \abs{v_k}^2$, with $c$ that does not depend on $k$~\cite{supp_mat}, $V(k) = \sum_p A(p, k ) A(k, p) \sim f_\alpha(k) / \lambda_\alpha(k) $ and $U(k) = V(-k)$. The behaviour of both integrals for $\ell\to\infty$ can be extracted by integrating $R_0(k)$ and $R_1(k)$ for $k\to 0$. In this limit, $f_\alpha(k)$,  and thus the single-particle energy $\lambda_\alpha(k)$, display a non-analytical scaling $f_\alpha(k)\sim\abs{k}^{\alpha-1}$~\cite{supp_mat}. 
For the first term in the r.h.s. of Eq.~(\ref{eqn_correlator_perturbation}) the latter behavior results in (details in Ref.~\cite{supp_mat})
\begin{equation}\label{eqn_GreenFunctionNoDisorder}
\braket{\Omega_0 	| a^\dag_j a_{j+\ell} |\Omega_0} \sim  
\begin{cases}
{1}/{\ell^{2-\alpha}}& \text{for } \alpha < 1 \\
{1}/{\ell^{2\alpha-1}}& \text{for } 1< \alpha < 2 \\
{1}/{\ell^{\alpha + 1}} &\text{for } \alpha > 2 
\end{cases}\end{equation}
which corresponds to the expected long-distance power-law decay of correlation functions for the homogeneous gapped superconductor with long-range pairing~\cite{Vodola2014, Vodola2016, Lepori2016, Maghrebi2016, Lepori2017}. Instead, for $R_1(k)$ the scaling of $f_\alpha(k)$ near $k\to 0$ implies
\begin{equation}
R_1(k) \sim \begin{cases}
k^{1-\alpha}    & \text{for } \alpha < 1 \\
k^{\alpha-1}   & \text{for }  \alpha > 1
\end{cases}
\end{equation}
which entails the following form of the disordered part of $C(\ell)$
\begin{equation}\label{eqn_correctionGreenFunction}
\braket{\delta\Omega_0 	| a^\dag_j a_{j+\ell} |  \delta\Omega_0}_W  
\sim
\begin{cases}
{W^2}/{\ell^{2-\alpha}}& \text{for } \alpha < 1 \\
{W^2}/{\ell^{\alpha}} &\text{for } \alpha > 1 
\end{cases}
\end{equation}
after the integration  of $R_1(k)$ in Eq.~\eqref{eqn_IntegralRandomCorr}~\cite{supp_mat}.

The discussion above demonstrates the following surprising results: (i) For $\alpha<1$ disorder is an irrelevant perturbation that does not modify the power of the algebraic decay of correlations, rather it affect its strength. (ii) For $\alpha>1$, the decay of correlations due to disorder is always algebraic, with an exponent that is smaller than for the homogeneous case with $W_j=0$. This implies that disorder enhances quasi-long-range order in these gapped models. (iii) For $\alpha \leq 2$ we find the duality relation $\gamma(\alpha) = \gamma(2 - \alpha)$ in the exponents of the algebraic decay.
This is reminiscent of the duality recently found for the decay exponent of the wave functions of long-range non-interacting spin models with positional disorder~\cite{Deng2018}. We come back to this point below.

The density-density correlation functions $G(j,\ell) = \left[\braket{n_{j} n_{j+\ell}} - \braket{n_{j} }\braket{n_{j+\ell}}\right]_W$ can also be obtained from the single-particle correlators $\braket{ a^\dag_{j} a_{j+\ell}}$ and $\braket{ a^\dag_{j} a^\dag_{j+\ell}}$ by means of the Wick theorem.  Examples of $G(\ell) $ are reported in~\cite{supp_mat}. Numerically we find that in the localized phases for model (I) when $\alpha<1$, $G(\ell)\sim 1/\ell^2$ while for both models $G(\ell)\sim 1/\ell^{2\alpha}$ when $\alpha>1$. The former behaviour with a decay exponent that does not depend on $\alpha$ is identical to that already observed in Refs.~\cite{Vodola2014,Vodola2016} in the absence of disorder. This underlines the irrelevance of disorder for $\alpha<1$. For $\alpha>1$, the decay is explained by considering the limit $\ell\to\infty$ of $\abs{C(\ell)}^2 \sim 1/\ell^{2\alpha}$ in Eq.~\eqref{eqn_correctionGreenFunction}.

\begin{figure}
\includegraphics[width=\columnwidth]{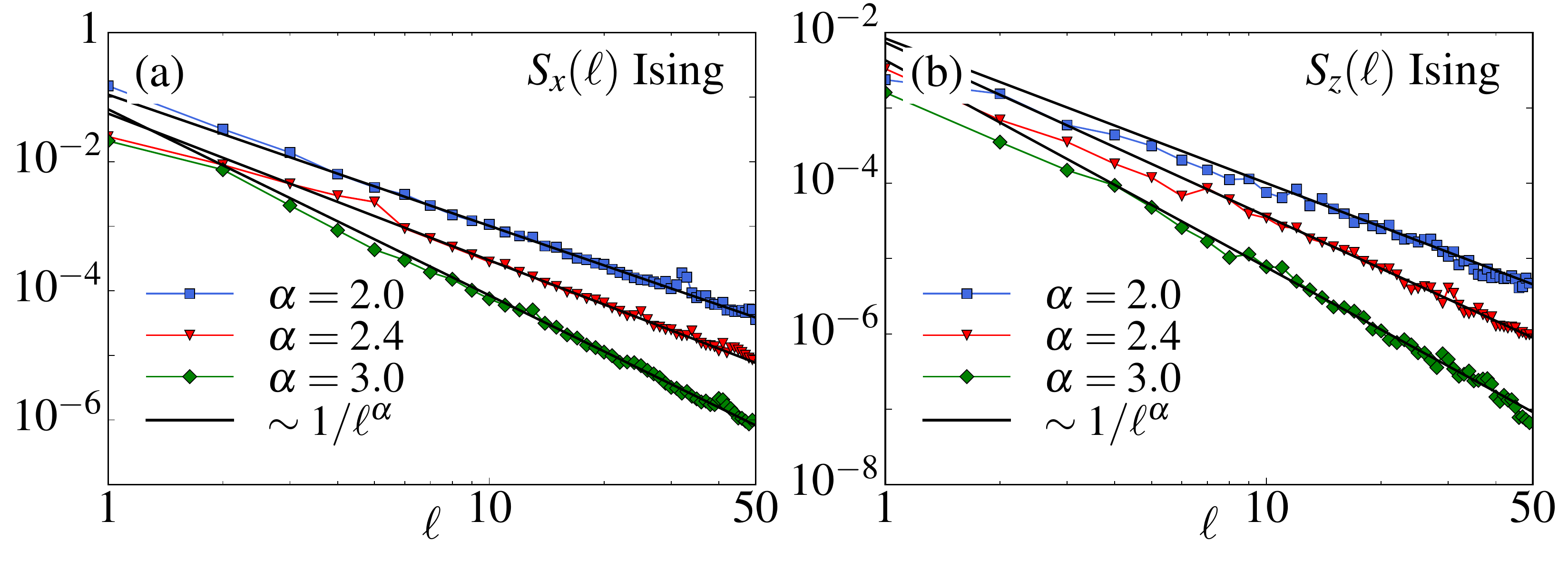}
\caption{(a) Correlation function $S_x(\ell)$ for the long-range Ising model with a random transverse field [$W=5\sin(\pi/5)$] and a constant interaction term ($B = 0$) for a system of $L=100$ spins and 50 disorder realizations. (b) Correlation function $S_z(\ell)$ for the long-range Ising model with a random interaction [$B = 5\sin(\pi/5)$] and a constant magnetic field ($W=0$). In both panels, the power-law tails are fit by the black lines scaling as $1/\ell^{\alpha}$.}\label{fig_CorrelationFunctionsIsing}
\end{figure}

For the random interacting long-range Ising model, we compute the spin-spin correlation functions  $S_\nu(\ell) := S_\nu(j_0,\ell)$ ($\nu=x,z$) within the MBL phase with $\alpha>1$, by using a DMRG alghoritm~\cite{White1992}. Here we choose $j_0 = L/10$. For the simulations, we use up to 400 local DMRG states, 16 sweeps and we average $S_{\nu}(\ell)$ over 100 disorder realizations. Strikingly, we find that $S_{\nu}(\ell)$ decays algebraically with $\ell$ as  $S_{\nu}(\ell)\sim \ell^{-\gamma}$ with an exponent that is consistent with $\gamma=\alpha$, in complete agreement with the discussion above for non-interacting theories. As an example, Fig.~\ref{fig_CorrelationFunctionsIsing}(a) shows $S_{x}(\ell)$ as a function of $\ell$ for different values of $\alpha$, $W=5\sin(\pi/5) \approx 2.93$ and $B = 0$, while Fig.~\ref{fig_CorrelationFunctionsIsing}(b) shows $S_{z}(\ell)$ as a function of $\ell$ for different values of $\alpha$, $W=0$ and $B = 5\sin(\pi/5)$. The corresponding fits (continuous lines) with $1/\ell^{\alpha}$ perfectly match the numerical results. 

The demonstration of quasi-long range order found in long-range couplings in the presence of disorder is a central result of this work. We argue that the fact that these results are found both for non-interacting and interacting models  strongly suggests the existence of a universal behavior due to long-range coupling.

\textit{Localization of wave functions.---} Numerical results on the decay of the single-particle wave functions  are obtained by considering the mean value $\Phi(\ell) = \sum_{q=1}^N\abs{g_{q,\ell-j_M}}/N$ where we average ${N} = L/4$ wave functions $g_{q,\ell}$ with lowest energies, shifted by the quantity $j_M$ that corresponds to the lattice site where $\abs{g_{q,\ell}}$ shows its maximum value. We average $\Phi(\ell)$ also over several disorder realizations (of the order of 500).

\begin{figure}
\includegraphics[width=\columnwidth]{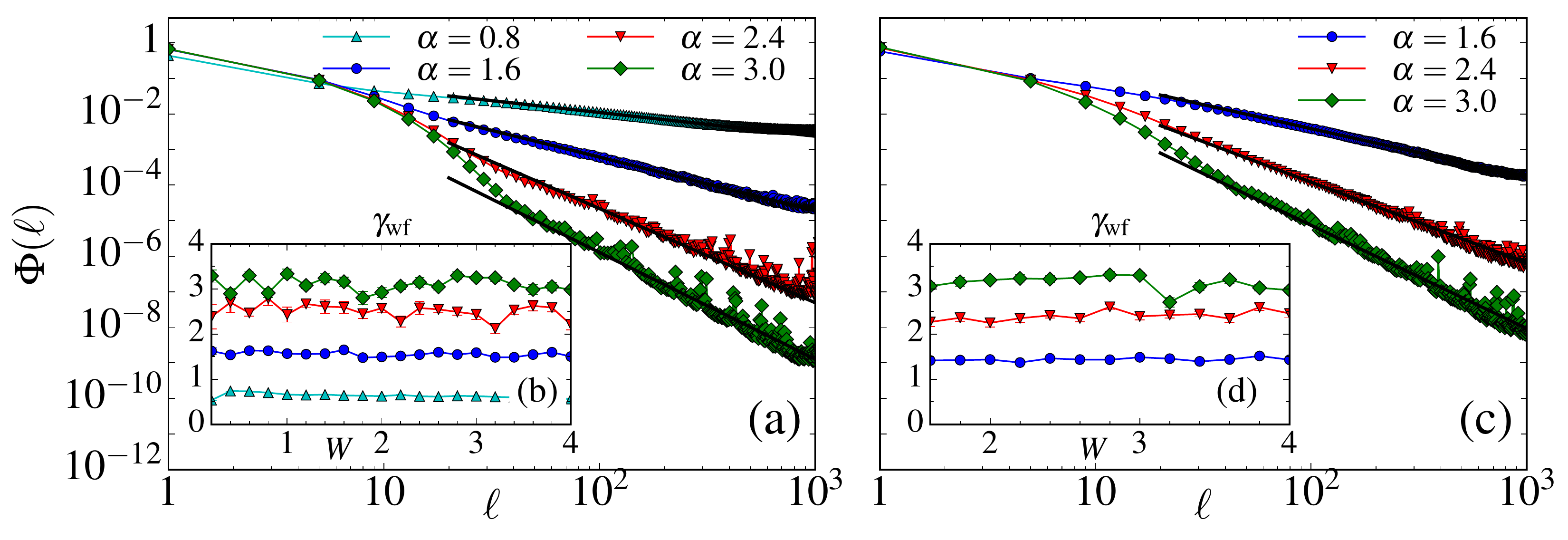} 
\caption{(a) Decay of the averaged wave function $\Phi(\ell)$ (absolute value, see text) of localized states for the model (I): If $\alpha>1$ we find an hybrid exponential and power-law behaviour. If $\alpha<1$ the exponential part is suppressed and only the power-law tail is visible. The black lines correspond to fit of the data scaling as $1/\ell^{\gamma_{\rm wf}}$. (b) Decay exponent $\gamma_{\rm wf}$ for the model (I) of the long-distance tail of $\Phi(\ell)$ as a function of $W$ for different values of $\alpha$. The decay exponent satisfies $\gamma_{\rm wf}\sim\alpha$ and does not show significance dependence on $W$. (c-d) Same as (a-b) but for the model (II) with random long-range pairing.}\label{fig_decaywavefunction}
\end{figure}

Figure~\ref{fig_decaywavefunction} shows typical results of the decay of $\Phi(\ell)$ as a function of the distance $\ell$ within the localized phases of models (I) and (II) of Eqs.~(\ref{eqn:kitaevHamiltonian}) [panels (a,b) and (c,d), respectively].

Remarkably, we find that the wave functions decay algebraically at long distances regardless of the strength $W$ of the disorder, mimicking the scaling of the correlation functions discussed above. However, for all $\alpha$, i.e. both  $\alpha> 1$ and $\alpha<1$, $\Phi(\ell)$ decays at large distances as $\Phi(\ell)\sim \ell^{-\gamma_{\rm wf}}$, with an exponent $\gamma_{\rm wf}$ consistent with $\gamma_{\rm wf}\sim \alpha$. This is different from the results of Ref.~\cite{Deng2018} with positional disorder, where for $\alpha<1$ one gets $\gamma_{\rm wf}\sim 2-\alpha$. For sufficiently large $\alpha>1$ this algebraic decay is preceded by an exponential decay at short distances, reminiscent of the exponentially localized states of short-range random Hamiltonians. 

In summary, we have demonstrated that interactions that decay as a power-law with distance induce an algebraic decay of correlation functions and wave functions both in non-interacting and interacting models in the presence of disorder. This is in stark contrast to results expected from short-range models, and generalises recent results for the decay of wave-functions in quadratic models. These results are of immediate interest for experiments with cold ions, molecule, Rydberg atoms and quantum emitters in cavity fields, to name a few. It is an exciting prospect to explore the properties of many-body quantum phases in the search of exotic transport phenomena with long-range interactions. 

P.~N.~ thanks L.~Benini for fruitful discussions. 
G.~P.~acknowledges support from ANR ``ERA-NET QuantERA'' - Projet ``RouTe'' and UdS via Labex NIE. 
E.~E.~is partially supported through the project ``QUANTUM'' by Istituto Nazionale di Fisica Nucleare (INFN) and through the project ``ALMAIDEA'' by University of Bologna.  The DMRG simulations were performed using the ITensor library~\cite{itensor}.

\pagebreak
\widetext    
\setcounter{equation}{0}
\makeatletter 
\renewcommand{\theequation}{S\@arabic\c@equation}
\makeatother

\setcounter{figure}{0}
\makeatletter 
\renewcommand{\thefigure}{S\@arabic\c@figure}
\renewcommand{\bibnumfmt}[1]{[S#1]}
\renewcommand{\citenumfont}[1]{S#1}
\makeatother

\onecolumngrid

\begin{center}
{\bf \large Algebraic Localization from Power-Law Interactions in Disordered Quantum Wires}\\[8pt]
\textsf{\textbf{ Supplemental Material}}\\[8pt]
T. Botzung,$^{1,2,3}$ D. Vodola,$^{4}$ P. Naldesi,$^{5}$ M. M\"uller,$^{4}$ E. Ercolessi,$^{2,3}$ G. Pupillo$^{1}$ \\
{\small
{$^1$}{\it University of Strasbourg, CNRS, ISIS (UMR 7006) and IPCMS (UMR 7504), 67000 Strasbourg, France} \\
{$^2$}{\it Dipartimento di Fisica e Astronomia dell'Universit\`a di Bologna, I-40127 Bologna, Italy} \\
{$^3$}{\it INFN, Sezione di Bologna, I-40127 Bologna, Italy} \\
{$^4$}{\it Department  of  Physics,  Swansea  University,  Singleton  Park,  Swansea  SA2  8PP,  United  Kingdom} \\
{$^5$}{\it Universit\'{e} Grenoble-Alpes, LPMMC, F-38000 Grenoble, France and CNRS, LPMMC, F-38000 Grenoble, France} \\
}
\end{center}

{\small
In this Supplemental Material we present details on the numerical simulations  and on the analytical derivations related to the $p$-wave superconducting random models that are not shown in the main text. Specifically, in Sec.~\ref{section_entropy} we analyze the scaling of the von Neumann entropy that gives further information on the localization properties of the wave functions of model (I). In Sec.~\ref{section_correlation} we give the details on the analytical computation of the decay of the single-particle correlation functions, and we show the decay exponents computed numerically. Finally, we show the behavior of the density-density correlation functions.
}

\section{Entanglement entropy}\label{section_entropy}
In this Section we give further insight on the localization properties of the states of Hamiltonians (I) by analysing the entanglement properties of their eigenmodes.

Measures of entanglement have been widely used to characterise the properties of ground states of many-body quantum systems~\cite{supp_Vidal2003} as well as to quantify the degree of localisation for ground- and excited states of disordered models~\cite{supp_Luitz2015}. A non-trivial measure of the rate of entanglement for a state $\ket{\phi}$ is the von Neumann entropy $S_\text{vN}(\phi,\ell) = -\Tr \rho_\ell \log_2 \rho_\ell$, where $\rho_\ell= \Tr_{L\setminus  \ell}\ket{\phi}\bra{\phi}$ is the reduced density matrix of the state $\ket{\phi}$ that contains $\ell$ sites of the entire lattice. $S_\text{vN}$ is known to follow an area-law scaling for localized states $\psi_\text{loc}$ [i.e. $S_\text{vN}(\psi_\text{loc},\ell)\sim \ell^{0}$], while for extended states $\psi_\text{ext}$ it it follows a volume law, e.g. it scales as $S_\text{vN}(\psi_\text{ext},\ell)\sim \ell$~\cite{supp_Bauer2013, supp_Li2015, supp_Geraedts2016}. In the following we compute $S_\text{vN}$ semi-analytically for a bipartition of the chain into two equal halves ($\ell=L/2$) for the excited states. An excited state of the Hamiltonians $H_\text{I,II}$ is defined by assigning a set of occupied modes $\mathbf{n}=\set{n_1,n_2,\ldots, n_L}$ with $n_q = 0,1$ and then creating single quasi-particles $\eta^\dag_q$ on the ground state $\ket{\Omega}$ if the mode $q$ is occupied 
\begin{equation}\label{eqn:ManyParticleStates}
\ket{\mathbf{n}} = \prod_{q=0}^{L-1} [{\eta_q^{\dagger}}]^{n_q} \ket{\Omega}.
\end{equation}

The two classes of excited states that we consider for computing the von Neumann entropy are given by 
\begin{gather}
\ket{\mathbf{n}_\nu} = \ket{\underbrace{0\dots0}_{\nu}\underbrace{11\dots11}_{L/4} 0\dots0}\label{eqn:ExcitedStates}\\ 
\ket{\mathbf{n}'_\nu} = \ket{\underbrace{0\dots0}_{\nu}\underbrace{1010\dots1010}_{L/4} 0\dots0} \label{eqn:ExcitedStates2}.
\end{gather}
We study the scaling of the von Neumann entropy as a function of the energy $e(\nu) = \sum_q n_q \Lambda_q$ and the system size $L$.

Following Ref.~\cite{supp_Peschel2003}, we compute the entropy of the excited states as a function of their energy and, by changing~$\nu$, we can explore the whole energy spectrum. This will provide a complete understanding of the different scalings of $S_\text{vN}$ with $L$ for high- and low-energy states. 

In order to compare the entropies of different eigenmodes, we first rescale the energies by introducing $\epsilon = [e(\nu)-E_\text{min}]/(E_\text{max}-E_\text{min})$, where $E_\text{min}$ ($E_\text{max}$) is the minimum (maximum) among the energies of the excited states of Eqs.~\eqref{eqn:ExcitedStates} or~\eqref{eqn:ExcitedStates2}. We then average the entropies, after binning them into groups of equal energy width.

\begin{figure}
\centering
\includegraphics[scale=0.35]{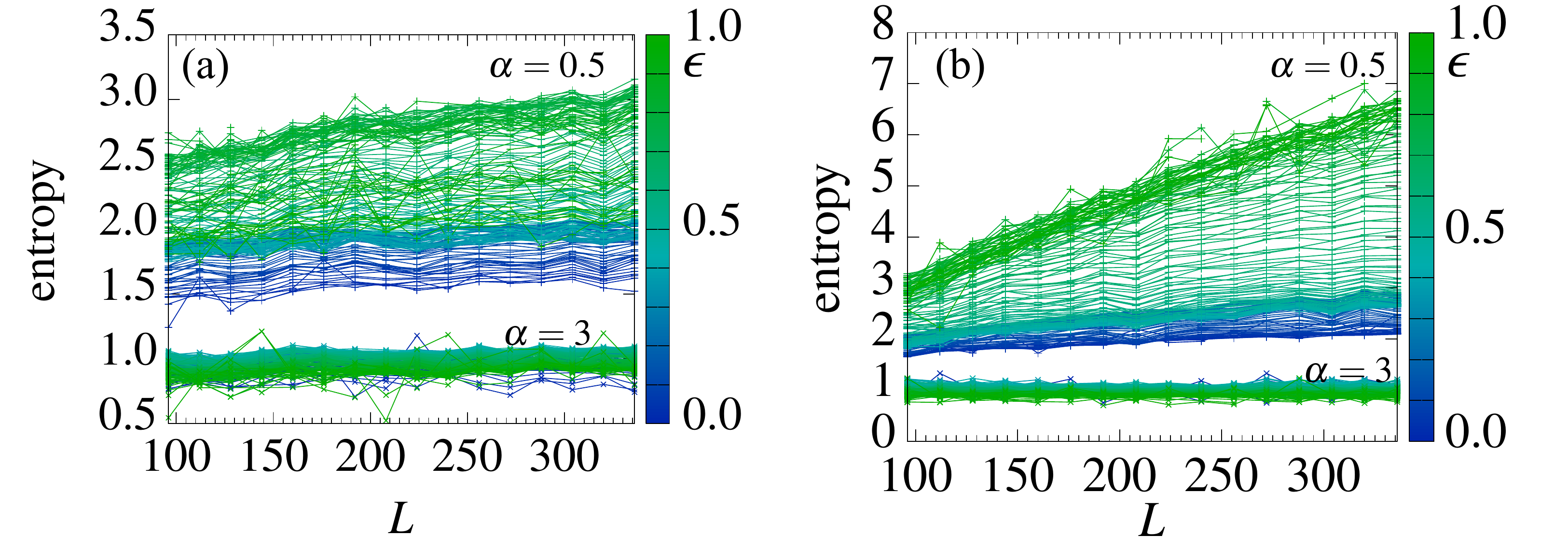}
\caption{von Neumann entropy $S_\text{vN}$ for the many-particle excited states of Hamiltonian (I) as a function of the system size $L$ for different energies $\epsilon$ and $W=1$: (a) states $\ket{\mathbf{n}}$ from Eq.~\eqref{eqn:ExcitedStates}, (b) states $\ket{\mathbf{n}'}$ from Eq.~\eqref{eqn:ExcitedStates2}. For $\alpha=3.0$, as all single-particle modes are localized,  the entropy of the states for all energies does not depend on the system size (i.e the von Neumann entropy satisfies an area law).   For $\alpha=0.5$, the scaling of the entanglement entropy depends on the energy of the excited states $\ket{\mathbf{n}}$ and $\ket{\mathbf{n}'}$: it goes from approximately constant for the low-energy states (depicted in blue), while it is found to follow a volume law (i.e., $S_\text{vN}\sim L$) for the high-energy ones (depicted with green lines). }\label{fig:EntropyI}
\end{figure}

In Fig.~\ref{fig:EntropyI} we show the entanglement entropy of the excited states of Hamiltonian (I) as a function of the system size $L$ for a given choice of $W = 1$ and for $\alpha=3$ and $\alpha=0.5$. Panel (a) shows the entropy for the excited states $\ket{\mathbf{n}}$ defined in Eq.~\eqref{eqn:ExcitedStates} while panel (b) shows the entropy for the excited states $\ket{\mathbf{n}'}$ defined in Eq.~\eqref{eqn:ExcitedStates2}. For $\alpha=3$, the entropy shows an area-law behavior (i.e., $S_\text{vN} \sim L^0$) for both the types of excited states at all energies. That behavior can be explained by the localisation of all single-particle modes. For $\alpha=0.5$, instead, the scaling of the entanglement entropy depends on the energy of the excited states $\ket{\mathbf{n}}$ and $\ket{\mathbf{n}'}$: it goes from approximately constant for the low-energy states (depicted in blue), while it is found to follow a volume law (i.e., $S_\text{vN}\sim L$) for the high-energy ones (depicted with green lines). We notice that the changing in the behavior of the entropy from area law to volume law is enhanced for the states~$\ket{\mathbf{n}'}$. This behaviour is compatible with the presence of a mobility edge for all $\alpha<1$ that separates localized low-energy states from extended high-energy ones. 

\section{Decay of correlation functions}\label{section_correlation}
In this Section we show how to compute the correlation function $C(j,i)=\braket{ a^\dag_j a_i}$ by perturbation theory. We will discuss only the model with random hopping, as the one with random long-range pairing can be treated similarly.

\subsection{Correlation functions - Perturbation theory}
We recall that the Hamiltonian $H_{\rm I}$ in Eq.~(1) of the main text is formed by two parts: 
\begin{equation}\label{supp_eqn_kitaevHamiltonian}
H_{{\rm I}}   =  H_0 + V_{{\rm I}}
\end{equation}
where
\begin{gather}
\label{supp_eqn_Hzero}
H_0 =  -t\sum_{j=1}^{L} \left(a^\dagger_j a_{j+1} + \mathrm{H.c.}\right)  + \mu \sum_{j=1}^{L} a^\dag_j a_j  +
\sum_{j,\ell} \frac{\Delta}{\ell^{\alpha}} \left( a_j a_{j+\ell}  + \mathrm{H.c.} \right), \\
\intertext{and}
\label{supp_eqn_randomHopping} V_{\rm I}  = -t \sum_{j=1}^{L} W_j \left(a^\dagger_j a_{j+1} + \mathrm{H.c.}\right).
\end{gather}

In order to compute the correlation function $C(j,i)=\braket{\Omega | a^\dag_j a_i|\Omega}$ on the ground state $\ket{\Omega}$ of $H_\text{I}$ in Eq.~\eqref{supp_eqn_kitaevHamiltonian}, we first find the first-order correction $\ket{\delta\Omega_0}$ to the ground state $\ket{\Omega_0}$ of $H_0$ by treating $V_\text{I}$ as a perturbation. 

The first-order correction $\ket{\delta\Omega_0}$ to the ground state $\ket{\Omega_0}$ of the Hamiltonian $H_0$ due to the perturbation $V_\text{I}$ is given by~\cite{supp_Sakurai2011}
\begin{equation}\label{supp_eqn_PerturbationTheory}
\ket{\delta\Omega_0} = \sum_{\mathbf{n}_0}\frac{\braket{ \mathbf{n}_0 | V_\text{I}| \Omega_0} }{E(\mathbf{n}_0)-E_0}\ket{\mathbf{n}_0}
\end{equation}
where, the quantities $E(\mathbf{n}_0)$ and $E_0$ are the energy of the states $\ket{\mathbf{n}_0}$ and of $\ket{\Omega_0}$, respectively and $\ket{\mathbf{n}_0}$ indicates an excited state of the homogeneous Hamiltonian $H_0$ that can be diagonalized via Fourier and Bogoliubov transformations as  
\begin{equation}\label{supp_eqn_HamiltonianBogoliubov}
H_0 = \sum_k \lambda_{\alpha}(k) \xi^\dag_k \xi_k.
\end{equation}
The ground state $\ket{\Omega_0}$ of $H_0$ is then the vacuum of all quasi-particles $\xi_k$. 

In Eq.~\eqref{supp_eqn_HamiltonianBogoliubov} we have defined the single-particle energy
\begin{equation}\label{supp_eqn_SingleParticleEnergy}
\lambda_{\alpha}(k) = [\left( \cos k - \mu \right)^2 + 4f_{\alpha}^2(k)]^{1/2}
\end{equation}
and the Bogolioubov quasi-particles $\xi_k$ that are related to the original fermionic operators $\tilde{a}_k$ in momentum space via 
\begin{equation}\label{supp_eqn_BogoliubovTransf}
\tilde{a}_k = v_k  \xi_k - u_k  \xi^\dag_{-k}
\end{equation}
with $v_k= \cos\varphi(k)$ and $u_k= \ii \sin\varphi(k)$ where $\tan 2\varphi(k) =  f_{\alpha}(k) / [\mu - \cos k]$ and $f_{\alpha}(k) = \sum_{\ell=1}^{L-1} \sin(k \ell)/\ell^\alpha$.  We notice that the functions $f_{\alpha}(k)$ when $L\to\infty$ become $f_{\alpha}(k) =  \left[\Li{\alpha}(\nepero^{\ii k})-\Li{\alpha}(\nepero^{-\ii k})\right]/(2\ii)$, with $\Li{\alpha}(z) = \sum_{j}z^j/j^\alpha$ a polylogarithm of order $\alpha$.

The excited states $\ket{\mathbf{n}_0}$ are defined by assigning a set of occupied modes $\mathbf{n}_0=\set{n_1,n_2,\ldots, n_L}$ with $n_q = 0,1$ and then creating single quasi-particles $\xi^\dag_q$ on the ground state $\ket{\Omega_0}$ if the mode $q$ is occupied 
\begin{equation}\label{eqn:ManyParticleStates}
\ket{\mathbf{n}_0} = \prod_{q=0}^{L-1} [{\xi_q^{\dagger}}]^{n_q} \ket{\Omega_0}.
\end{equation}

The first-order correction $\ket{\delta\Omega_0}$ can now be obtained from  Eq.~\eqref{supp_eqn_PerturbationTheory} that gives

\begin{equation}\label{eqn:PerturbedGSHopping}
\ket{\Omega} =\ket{\Omega_0} + \ket{\delta\Omega_0}=\ket{\Omega_0} -  \sum_{kk'} {J}_{k,k'} A(k,k')  \xi^\dag_k \xi^\dag_{k'}\ket{\Omega_0},
\end{equation}
where we have defined ${J}_{k,k'} = -\sum_j e^{\ii (k-k') j} W_j/L$ and $A(k,k') = 2(\nepero^{\ii k} + \nepero^{-\ii k'}) v_k u^*_{k'} / [\lambda(k)+\lambda(k')]$.

On a single disorder realization the correlation function $\braket{\Omega | a^\dag_j a_i | \Omega }$ takes the form
\begin{equation}\label{supp_eqn_AllTermCorrelation} % \bra{\Omega_0} + \bra{\delta\Omega_0} \ket{\Omega_0} + \ket{\delta\Omega_0}
\braket{\Omega | a^\dag_j a_i | \Omega } =  
\braket{\Omega_0    | a^\dag_j a_i |      \Omega_0} +
\braket{\delta\Omega_0  | a^\dag_j a_i |  \Omega_0} +
\braket{\Omega_0 		| a^\dag_j a_i |  \delta\Omega_0} +
\braket{\delta\Omega_0 	| a^\dag_j a_i |  \delta\Omega_0}.
\end{equation}
If we now average Eq.~\eqref{supp_eqn_AllTermCorrelation} over many disorder realizations,  the cross terms $\braket{\delta\Omega_0  | a^\dag_j a_i |  \Omega_0}$ and $\braket{\Omega_0 		| a^\dag_j a_i |  \delta\Omega_0}$ vanish as, due to the correction  $\ket{\delta\Omega_0}$, only one random term $W_j$ (that has mean value zero) appears in them. Therefore we get
\begin{equation}\label{supp_eqn_AveragedCorrelation}
\braket{\Omega | a^\dag_j a_i | \Omega }_W =  
\braket{\Omega_0    | a^\dag_j a_i |      \Omega_0} +
\braket{\delta\Omega_0 	| a^\dag_j a_i |  \delta\Omega_0}_W.
\end{equation}
The first term of the r.h.s.\ of Eq.~\eqref{supp_eqn_AveragedCorrelation} corresponds to the correlator for a homogenous translationally-invariant system. By rewriting $a^\dag_j$  and $a_i$ in momentum space and by using Eq.~\eqref{supp_eqn_BogoliubovTransf} recalling that $\xi_k\ket{\Omega_0} = 0$ we obtain
\begin{equation}\label{supp_eqn_CleanCorrelator}
C_0(\ell) :=\braket{\Omega_0    | a^\dag_j a_i | \Omega_0} = \frac{1}{L}\sum_{k} \nepero^{\ii k \ell} R_0(k)
\end{equation}
where $\ell = j-i$ and $R_0(k) = \abs{u_k}^2$.

In the second term of the r.h.s.\ of Eq.~\eqref{supp_eqn_AveragedCorrelation}, as we are averaging on the disorder configurations, we can expect that the disorder average $\braket{\delta\Omega_0 | a^\dag_j a_i | \delta\Omega_0}_W$  will be translationally invariant, i.e.\@\xspace it will depend on the relative distance $\ell=j - i$ while the terms that depend on $i$ and $j$ separately will average out to zero (see \S6.5 in Ref.~\cite{supp_Altland2006} or \S12.3 in Ref.~\cite{supp_Bruus2004}). By keeping only the terms that depend on $\ell$, after rewriting $a^\dag_j$  and $a_i$ in momentum space and using again Eq.~\eqref{supp_eqn_BogoliubovTransf} recalling that $\xi_k\ket{\Omega_0} = 0$, the second term becomes
\begin{equation}\label{supp_eqn_DisorderCorrelator}
C_1(\ell) := \braket{\delta\Omega_0 	| a^\dag_j a_i |  \delta\Omega_0}_W = \frac{W^2}{3L}\sum_{k } \nepero^{\ii k \ell} R_1(k)
\end{equation}
where
\begin{align}
\label{supp_eqn_R1_def} R_1(k)&= c \abs{u_k}^2 + U(k) \abs{u_k}^2 - V(k) \abs{v_k}^2,  \phantom{\sum_p}\\ 
\label{supp_eqn_c_def} c &= \sum_p A(p, p)^2  -\sum_{p_1 p_2}  A(p_1, p_2) A(p_2,p_1), \\
\label{supp_eqn_U_def} U(k) &= 2\sum_p A(p, -k ) A(-k, p) = -\frac{f_\alpha(k)}{\lambda_{\alpha}(k)}\sum_p \frac{2 + 2 \cos(p-k)}{(\lambda_{\alpha}(k) + \lambda_{\alpha}(p))^2}\frac{f_\alpha(p)}{\lambda_{\alpha}(p)}, \\
\label{supp_eqn_V_def} V(k) &= 2\sum_p A(p, k ) A(k, p) = \frac{f_\alpha(k)}{\lambda_{\alpha}(k)}\sum_p \frac{2 + 2 \cos(p+k)}{(\lambda_{\alpha}(k) + \lambda_{\alpha}(p))^2}\frac{f_\alpha(p)}{\lambda_{\alpha}(p)}.
\end{align}
We note that the quantity $c$ does not depend on $k$.

\subsection{Correlation functions - Asymptotic behavior}
\begin{figure}
\begin{tikzpicture}
\def\bigradius{3}
\def\eps{0.12}
\draw [help lines,->] (-1.25*\bigradius, 0) -- (1.25*\bigradius,0);
\draw [help lines,->] (0, -0.25*\bigradius) -- (0, 1.25*\bigradius);
\draw [line width=0.75pt,   decoration={markings,  mark=at position 0.5 with {\arrow[line width=1.2pt]{>}}},   postaction={decorate}]  (\eps,0) -- (\bigradius,0) ;
\draw [line width=0.75pt,   decoration={markings,  mark=at position 0.5 with {\arrow[line width=1.2pt]{>}}},   postaction={decorate}]  (\bigradius,0) arc (0:88:\bigradius);  
\draw [line width=0.75pt,   decoration={markings,  mark=at position 0.55 with {\arrow[line width=1.2pt]{<}}},   postaction={decorate}]  (\eps,0) -- (\eps,\bigradius);
\draw [line width=0.75pt,   decoration={markings,  mark=at position 0.5 with {\arrow[line width=1.2pt]{<}}},   postaction={decorate}]  (-\eps,0) -- (-\bigradius,0) ;
\draw [line width=0.75pt,   decoration={markings,  mark=at position 0.5 with {\arrow[line width=1.2pt]{<}}},   postaction={decorate}]  (-\bigradius,0) arc (180:92:\bigradius);  
\draw [line width=0.75pt,   decoration={markings,  mark=at position 0.45 with {\arrow[line width=1.2pt]{>}}},   postaction={decorate}]  (-\eps,0) -- (-\eps,\bigradius);
\fill[red] (0,0) circle[radius=1.4pt];
\draw [line width=0.75pt,red, label=$x$] (0,0)  -- (0,1.1*\bigradius);
\node [above] at (110:0.5*\bigradius) {${s}_{-}$};
\node [above] at (150:1.05*\bigradius) {$\Gamma_-$};
\node [above] at (50:\bigradius) {$\Gamma_+$};
\node [below] at (70:0.5*\bigradius) {${s}_{+}$};
\node [below] at (0,0) {0};
\node [below] at (1*\bigradius,0) {$r$};
\node [below] at (1.9*\eps,0) {$\epsilon$};
\node [below] at (1.3*\bigradius,0) {$k=\mathrm{Re\,}z$};
\node [left] at (0,1.2*\bigradius) {$\mathrm{Im\,}z$};
%\node at (-1.2*\bigradius, 1.2*\bigradius) {(a)};
\end{tikzpicture}
\caption{Integration contour for evaluating the asymptotic behaviors of the correlators $C_0(\ell)$ in Eq.~\eqref{supp_eq_C0Integral} and $C_1(\ell)$ in Eq.~\eqref{supp_correlator_C1_integral}.}\label{supp_fig_contour}
\end{figure}
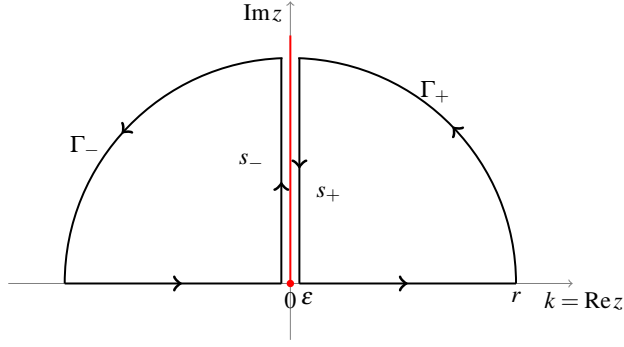

In this Section we show how the two correlators $C_0(\ell)$ and $C_1(\ell)$ behave asymptotically for $\ell\to\infty$.

%apply standard techniques for the to Fourier transformation~\cite{supp_Ablowitz2003}.

%The asymptotic behavior of $C_0(\ell)$ and $C_1(\ell)$ for $\ell\to \infty$ can be computed by 

Let us consider $C_0(\ell)$ in Eq.~\eqref{supp_eqn_CleanCorrelator} first. In the limit $L\to\infty$ we can replace the summation with an integral
\begin{equation}
C_0(\ell) = \frac{1}{2\pi}\int_{-\infty}^{\infty} \de k \ \nepero^{\ii k \ell} R_0(k).
\end{equation}

The asymptotic behavior of $C_0(\ell)$ for $\ell\to\infty$ can be computed by considering the integrals $I_0^+$ and  $I_0^-$ on the complex plane in Fig.~\ref{supp_fig_contour} that are
\begin{gather}
I_0^+ =  \frac{1}{2\pi}\int_{s_+} \de z \ \nepero^{\ii z \ell} R_0(z) +  \frac{1}{2\pi}\int_{\Gamma_+} \de z\ \nepero^{\ii z \ell} R_0(z) + \frac{1}{2\pi}\int_{0}^{\infty} \de k \ \nepero^{\ii k \ell} R_0(k) \\
I_0^- =  \frac{1}{2\pi}\int_{s_-} \de z \ \nepero^{\ii z \ell} R_0(z) +  \frac{1}{2\pi}\int_{\Gamma_-} \de z \ \nepero^{\ii z \ell} R_0(z) + \frac{1}{2\pi}\int_{-\infty}^{0} \de k \ \nepero^{\ii k \ell} R_0(k)
\end{gather}
where we have chosen to put the branch cut of the complex logarithm [see the expansion of the polylogarithm in Eq.~\eqref{supp_eqn_polylogarithm}] on the imaginary positive axis.

By sending the radius $r$ of the circles $\Gamma_\pm$ to infinity and by neglecting possible residues inside the integration contour that will contribute only with exponential decaying terms we have
\begin{equation}\label{supp_eq_C0Integral}
\begin{split}
C_0(\ell) & = -  \frac{1}{2\pi}\int_{s_+} \de z \ \nepero^{\ii z \ell} R_0(z) - \frac{1}{2\pi}\int_{s_-} \de z \ \nepero^{\ii z \ell} R_0(z) \\
& =  \frac{\ii}{2\pi}\int_0^\infty  \de y\ \nepero^{- y\ell} R_0(\epsilon + \ii y) -  \frac{\ii}{2\pi}\int_{-\infty}^{0}  \de y\ \nepero^{- y\ell} R_0(- \epsilon + \ii y) \\
& =  \frac{1}{\pi}\int_0^\infty  \de y\ \nepero^{- y\ell} \Im R_0(\ii y)
\end{split}
\end{equation}
where on the lines $s_{\pm}$ the complex variable is $z=\pm\epsilon + \ii y$ with $\epsilon$ a small positive parameter that we send to zero.

We are able now to evaluate the asymptotic behavior of $C_0(\ell)$ by computing the $y\to 0$ part of $\Im [R_0(\ii y)]$ and then integrating the last equality in Eq.~\eqref{supp_eq_C0Integral}. This is done by recalling that the polylogarithm admits the series expansion~\cite{supp_Abramowitz1964, supp_Olver2010} for a general complex number $z$ as
\begin{equation}\label{supp_eqn_polylogarithm}
\operatorname{Li}_\alpha(z) = \Gamma(1 - \alpha) \left(\ln \frac{1}{z} \right)^{\alpha-1} + \sum_{n=0}^\infty \zeta(\alpha-n)\frac{(\ln z)^n }{n!}
\end{equation}
that makes them non-analytical due to the presence of the complex logarithm and the power-law. In Eq.~\eqref{supp_eqn_polylogarithm}, $\Gamma(x)$ and $\zeta(x)$ are the Euler gamma function and the Riemann zeta function, respectively.

By using the series expansion of the polylogarithms from Eq.~\eqref{supp_eqn_polylogarithm} that yields
\begin{equation}
\Li{\alpha}(\nepero^{-y}) - \Li{\alpha}(\nepero^{y})   = \Gamma(1-\alpha) \left(1 + \nepero^{\ii \pi \alpha}  \right) y^{\alpha-1}  - 2 \sum_{n\text{ odd}}^\infty \frac{\zeta(\alpha-n)}{n!} y^{n} 
\end{equation}
we can obtain the function $R_0(\ii y)$ on the imaginary axis: 
\begin{equation}
R_0(\ii y) = \frac{\mu - \cosh y}{2\lambda_\alpha(\ii y)} \sim \frac{\mu-1}{2\sqrt{(\mu-1)^2-\Gamma^2(1-\alpha)(\nepero^{\ii \pi \alpha}+1)^2 y^{2\alpha-2} -4 \Gamma(1-\alpha)(\nepero^{\ii \pi \alpha}+1)\zeta(\alpha-1) y^\alpha}}.
\end{equation}
The previous equation in the limit $y\to 0$ gives
\begin{equation}
\Im R_0(\ii y) = \begin{cases}
y^{1-\alpha}    & \text{for } \alpha < 1 \\
y^{2\alpha-2}   & \text{for } 1< \alpha < 2 \\
y^{\alpha}  &\text{for } \alpha > 2 
\end{cases}
\end{equation}
and, after performing the last integral in Eq.~\eqref{supp_eq_C0Integral}, the asymptotic behavior of $C_0$ turns out to be
\begin{equation}
C_0(\ell) \sim
\begin{cases}
{1}/{\ell^{2-\alpha}}& \text{for } \alpha < 1 \\
{1}/{\ell^{2\alpha-1}}& \text{for } 1< \alpha < 2 \\
{1}/{\ell^{\alpha + 1}} &\text{for } \alpha > 2 .
\end{cases}\label{supp_CorrelatorCleanSystem}
\end{equation}

For the correlator $C_1(\ell)$ in Eq.~\eqref{supp_eqn_DisorderCorrelator} we can use the same contour in Fig.~\ref{supp_fig_contour} and get
\begin{equation}\label{supp_correlator_C1_integral}
C_1(\ell) = \frac{W^2}{3\pi}\int_0^\infty  \de y\ \nepero^{- y\ell} \Im R_1(\ii y).
\end{equation}

For the asymptotic behaviour of $C_1(\ell)$, we need again the  $y\to 0$ part of $R_1(\ii y)$.
Let us start by noting that from Eqs.~\eqref{supp_eqn_U_def} and~\eqref{supp_eqn_V_def} the $y\to 0$ part of both $U(\ii y)$ and $V(\ii y)$  is given by
\begin{equation}
\Im [U(\ii y) \abs{u_{\ii y}}^2] \sim \Im [V(\ii y) \abs{v_{\ii y}}^2] \sim \Im \frac{f_\alpha(\ii y)(\mu - \cosh y)}{\lambda^2_\alpha(\ii y)}\sim \begin{cases}  y^{1-\alpha}  &\text{for }  \alpha < 1 \\ 
y^{\alpha-1}  &\text{for }  \alpha > 1.
\end{cases}
\end{equation}
The previous equation, by considering also the contribution coming from $c \abs{u_{\ii y}}^2$ [see Eq.~\eqref{supp_eqn_R1_def}], gives
\begin{equation}
\Im R_1(\ii y) \sim \begin{cases}
y^{1-\alpha}    & \text{for } \alpha < 1 \\
y^{\alpha-1}  &\text{for } \alpha > 1
\end{cases}
\end{equation}
and after integrating Eq.~\eqref{supp_correlator_C1_integral}, we finally get the correlator
\begin{equation}
C_1(\ell) = \begin{cases}
{W^2}/{\ell^{2-\alpha}}& \text{for } \alpha < 1 \\
{W^2}/{\ell^{\alpha}} &\text{for } \alpha > 1.
\end{cases}\label{supp_CorrelatorFinal}
\end{equation}

\begin{figure}
\includegraphics[scale=0.4]{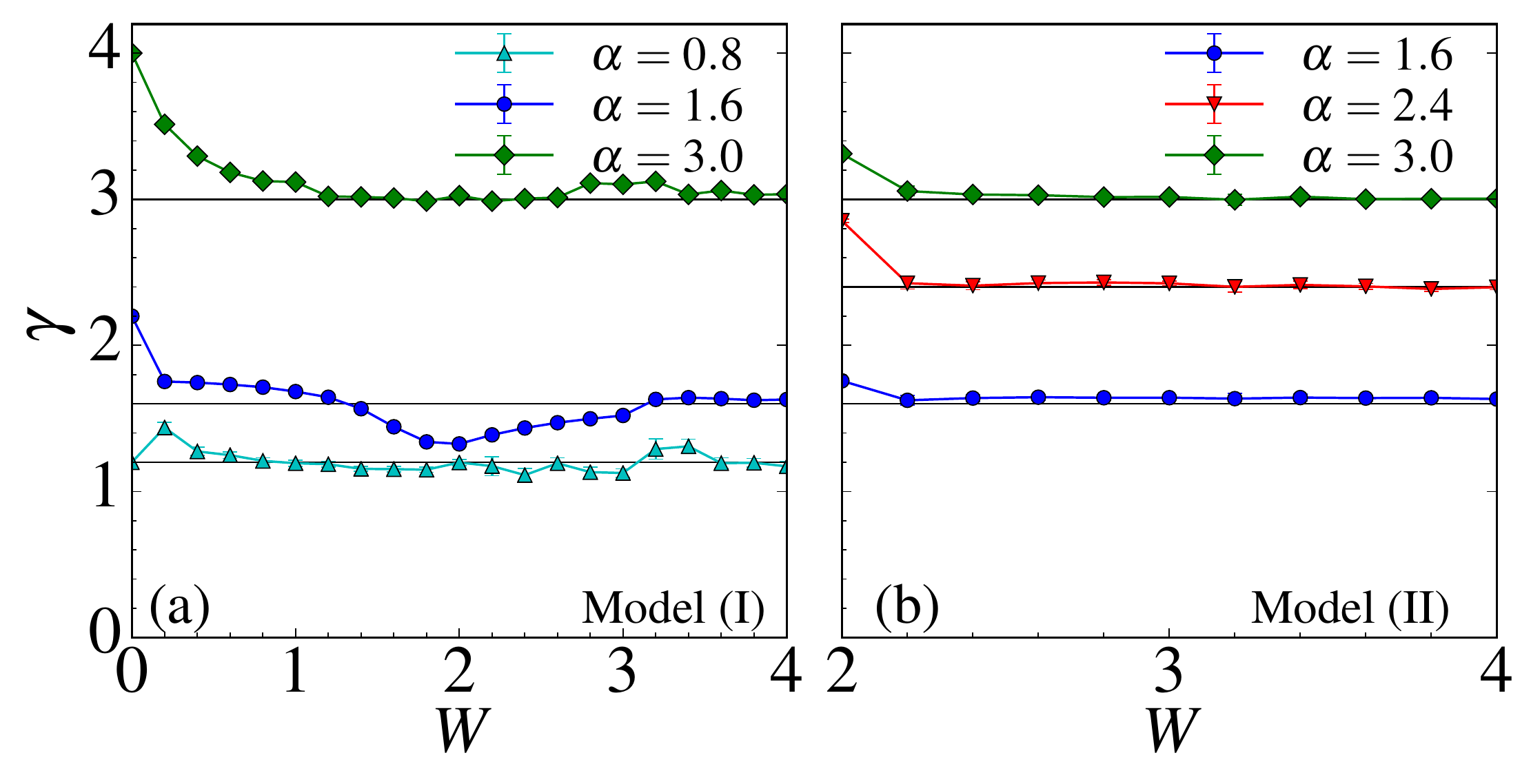}
\caption{(a) Decay exponent $\gamma$ of the long-distance tail of the correlation function $C(\ell)$ for the model (I) as a function of $W$ and for $\alpha = 0.8$ (cyan triangles), $\alpha = 1.6$ (blue circles), $\alpha = 3.0$ (green diamonds). If $W>0$, the decay exponent satisfies $\gamma \sim \alpha$ for $\alpha>1$ and $\gamma\sim 2-\alpha$ for $\alpha<1$ and it does not show significance dependence on $W$. These data are obtained by computing  the correlation function $C(\ell)$ numerically from the full random Hamiltonian in Eq.~\eqref{supp_eqn_kitaevHamiltonian} and then by fitting the long-range decaying tail of $C(\ell)$  with $1/\ell^\gamma$. The black lines represent the expected exponents: $\gamma=1.2$ for $\alpha=0.8$,  $\gamma=1.6$ for $\alpha=1.6$,  $\gamma=3.0$ for $\alpha=3.0$. (b) Same as panel (a) but for the localized phase (for $W\gtrsim 2$) of model (II).}\label{fig_decay_exponent_correlator}
\end{figure}

The asymptotic behavior coming from Eqs.~\eqref{supp_CorrelatorCleanSystem}, \eqref{supp_CorrelatorFinal} can be checked by computing the correlator $C(\ell)$ numerically as reported in Fig.~2(a) of the main text. Remarkably, the values of the decay exponents of the power-law tails do not depend on the disorder strength $W$ as shown in Fig.~\ref{fig_decay_exponent_correlator} where we plot the decay exponents of $C(\ell)$ as a function of $W$ for different values of $\alpha$. For completeness we show also the decay exponent of the correlation function $C(\ell)$ for the model (II) with random long-range pairing.

\subsection{Density-density correlation function}
\begin{figure}
\includegraphics[scale=0.35]{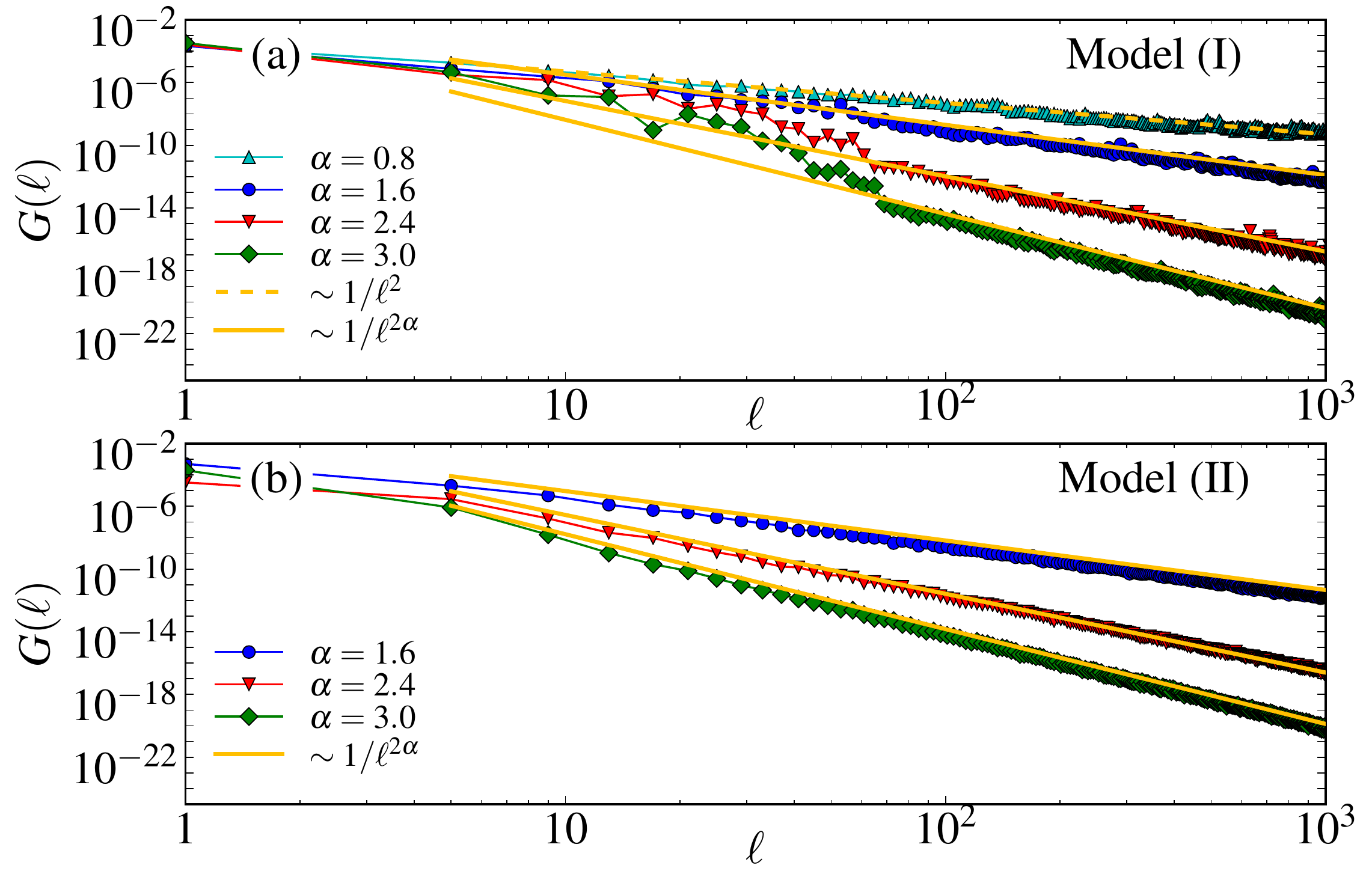}
\caption{(a) Density-density correlation function $G(\ell)$ for the model~(I) as a function of the lattice site $\ell$ for different values of $\alpha$ and for $W=5$, $L=2000$ and 400 disorder realizations. The power-law tails are fit by the yellow lines scaling as $1/\ell^{2}$ (dashed) and $1/\ell^{2\alpha}$ (solid). (b) Same as panel (a) but for the model (II).}\label{fig_DensDensCorrelationFunctions}
\end{figure}

From the single-particle correlators $\braket{ a^\dag_{j} a_{j+\ell}}$ and $\braket{ a^\dag_{j} a^\dag_{j+\ell}}$,  by means of the Wick theorem, we computed also the density-density correlation functions $G(j,\ell) = \left[\braket{n_{j} n_{j+\ell}} - \braket{n_{j} }\braket{n_{j+\ell}}\right]_W = [ \abs{\braket{a_{j} a_{j+\ell}}}^2 - \abs{\braket{a^\dag_{j} a_{j+\ell}}}^2]_W$.

Examples of $G(\ell) = G(j_0,\ell)$ with $j_0=L/4$  are shown in Fig.~\ref{fig_DensDensCorrelationFunctions} for a system of $L=2000$ sites and for a disorder strength $W=5$. Numerically we find that in the localized phases for model (I) when $\alpha<1$, $G(\ell)\sim 1/\ell^2$ while for both models $G(\ell)\sim 1/\ell^{2\alpha}$ when $\alpha>1$. The first behaviour with a decay exponent that does not depend on $\alpha$ has been already observed in Refs.~\cite{supp_Vodola2014,supp_Vodola2016}, while the second can be explained by looking at the $\ell\to\infty$ scaling of $\abs{C(\ell)}^2 \sim 1/\ell^{2\alpha}$ in Eq.~\eqref{supp_CorrelatorFinal}.

\clearemail

\end{document}